# NEW TASKS AND NEW CODES FOR RFQ BEAM SIMULATION

Boris Bondarev, Alexander Durkin, Stanislav Vinogradov, Igor Shumakov
*Moscow Radiotechnical Institute RAS*
132, Warshavskoe Shosse, Moscow, Russia, 113519
lidos@aha.ru

## INTRODUCTION

Proton linear accelerator is the base Accelerator Driven Power System (ADS). Such ADS are dedicated to various purposes: weapon plutonium conversion, "energy amplifier", transmutation of radionuclear wastes etc. Solution of these tasks requires proton beams with energy 1 GeV and average current up to 30 mA. At the moment there are no problems of fundamental nature in such linac construction. The main problems have economic and technical aspects.

Problems of CW linac will be demonstrated on the base beam dynamics requirements. New code package LIDOS.RFQ.Designer makes possible to simulate beam dynamics in RF fields of real vane shape (including gaps between RFQ section) as well as to determine channel parameters tolerances for reliable operation..

## 1. NEW UNIQUE CODES LIDOS.RFQ.DESIGNER

New criteria arise when high-current CW linacs are considered. The main requirements for such linacs are maximal RF intensity reduction very small beam losses. In such cases the traditional algorithms of RFQ designing can lead to undesired versions of the RFQ linac.

Minimization of CW beam losses in RFQ linac places more stringent requirements upon beam perturbations. Instrumental errors in vane manufacturing, installation and adjustment are sources of such perturbations. Even with very small cell parameter deviations the potential of such perturbations is high enough both for beam transmission reduction and beam quality degradation. A reason enough to such statement is provided by the fact that trajectories even for "ideal" (without perturbations) RFQ channel are spaced in the immediate vicinity of vane surfaces.

Gaps between RFQ sections are another sources of perturbations. The gaps divided RFQ channel into several sections are dictated by parasitic mode suppression. Focusing RF field reduction inside gap and in the adjacent area leads to beam mismatching, growths of beam size and emittance. Accelerating field change leads to local reduction of accelerating efficiency and to appearance of beam coherent phase oscillations. In order to avoid additional gap it is necessary to "insert" a gap not as additional channel segment but as a substitution of regular channel segment. By other words a regular RFQ cell is substitute by "cell" with gap. In this case phase length of cell is distinguished from its nominal value only by small amount (caused by difference in RF field amplitudes).

New concepts for parameter choices based on optimization methods and taken into account real RF fields are incorporated in the new multilevel codes [1,2]. There are two main features: a maximum of scientific visualization for each calculation step and the possibility to cut off undesired linac versions long before the time-consuming calculations start. The package contains codes with three levels of mathematical model complexity.

The **first-level** codes make only a preliminary choice of the main parameter arrays on the basis of a simplified physical model. These codes are richly supplied with visual information that helps to find the best linac version quickly. Separate algorithm branch allows using output parameter table obtained by PARMTEQ codes as initial information

The **second-level** codes are used for channel data calculations with the real shape of the RFQ vanes and real RF fields. Information from the first level codes is used here as input data.

The **third-level** codes are based on information from the first and second level codes and on complex PIC-models that are needed for a correct beam simulation in the chosen channel version.

The package gives users additional possibilities in comparison with existing codes (PARMTEQ and so on). In the frame of new codes it is possible to:
- calculate RF field mesh values taking into account vane real shape and gap inserting,
- simulate beam motion in calculated RF fields taking into account field perturbations caused by instrumental errors in vane manufacturing, installation and adjustment (including cases with symmetry violation),
- make statistical analysis of output parameter degradation.

## 2. CALCULATIONS IN FAVOR OF IPHI RFQ AS DEMONSTRATION OF CODE POWER

The code tools described above were used for RFQ designing in favor of IPHI Project (CEA, France) [3].

Below the main results are presented to demonstrate new code power.

The following parameters were preset:

| | |
|---|---|
| Accelerating Particles | Protons |
| Input Energy | 0.095 MeV |
| Output Energy | 5 MeV |
| Beam Current | 100 mA |
| Beam Emittance | 0.15π cm·mrad |
| Operating Frequency | 352 MHz |

There were some restrictions and requirements:

| | |
|---|---|
| Kilpatrick Factor | no more than 1.6 |
| Maximum RF Field Intensity | less than 29.5-30 MV/m |
| Length | about 8 m |
| Beam Transmission | no less then 90% |
| Power of Lost Particles | no more than 1.5 kW |
| Minimum of Lost Particles With Energy | more then 3 MeV |

## 3. POSSIBLE TYPES OF PERTURBATION

Random errors independently arising in each cell are the most danger. In this case there is a possibility of unfavorable realization with each next cell amplifying disturbances of all previous ones.

Analytical theory for transverse size statistic estimation as well as for emittance growth by the action of random perturbation was generated early [4,5]. The same methods regarding to RFQ channel are given below.

### 3.1. Random Deviation of Focusing Field Gradient

Let us consider that errors of focusing field gradient are caused by random deviations of cell aperture radius. In this case probability distribution function has a form

$$P(Q_T) = 1 - \exp(-(\ln Q_T)^2/(2N\Delta^2))$$

Where $P(Q_T)$ is probability of effective emittance growth no more then times $Q_T$ times.

It follows, that the effective emittance growth coefficient will not be exceed the $a$ limit with the probability $p$, if tolerances are defined by the equation

$$N\Delta^2 = -\frac{(\ln a)^2}{2\ln(1-p)}$$

The next relation can be used for practical estimates

$$\Delta = \frac{eU\lambda}{8W_0 gE} \frac{x_{max}^2 + x_{min}^2}{2R_0^2}$$

where $W_0$ is ion rest energy, $E$ is beam emittance, $x_{max}$ and $x_{min}$ are maximal and minimal values of matched beam envelope, $U$ is intervane voltage, $\lambda$ is wave length of RF field.

### 3.2. Displacement of the Focusing Channel Axis.

Random displacements of focusing elements with $s$ as rms value give rise to rms-displacement D of the beam center described by the relation

$$\Delta^2 = G^2 N s^2$$

where

$$G = \frac{eU\lambda}{2W_0 gE} \frac{x_{max} + x_{min}}{2R_0^2}$$

In this case, the transverse motion amplitude is a random value distributed by the following law

$$\mathbf{F}(x) = 1 - \exp(-x^2/\Delta^2)$$

The Monte-Carlo simulations were performed with the aim to estimate how possible deviations of accelerating and focusing fields can be differed the output beam parameters. Change in averaged radius position for each cell was considered as perturbation factor. It was assumed that averaged radius inside cell varies linearly. The random value distributed uniformly inside [-δr, δr] was added to top value of cell averaged radius. Random values for different cells were chosen independently. Beam simulations were performed for δr =10 ìm and δr =25 ìm. The code package described above was made statistical data processing. Beam energy losses, phase width, momentum spread, transverse size, beam center displacement, beam rms and total emittances were calculated as integral characteristics. These values were integrated over all particles inside one RF period. Phase space area occupied by all macroparticles was nominated as total emittance.

Random displacements of RFQ section ends were study separately. In this case RFQ axis was presented as four straight-line segments with displacement amplitude är = 100 μm.

The number of random realizations is 50 for each type of errors. In the next tables more intrusting results for three types of random errors are presented.

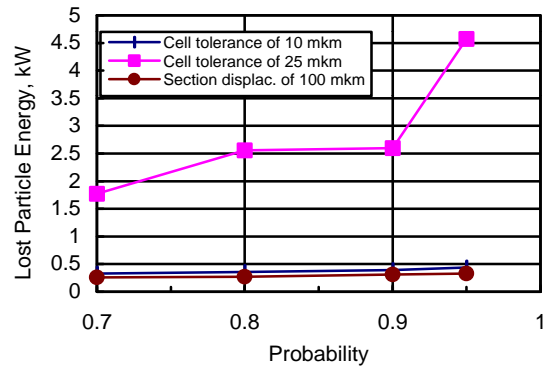

Fig.1. Lost Particle Total Energy (In Ideal Channel Lost Particle Total Energy Equals 0.134 kW)

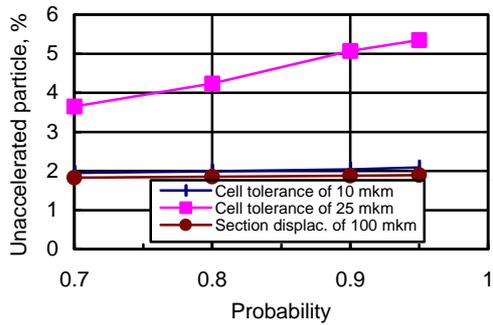

Fig.2. Unaccelerated Particles (in %) vs probability. Ideal Value equals 1.68%.

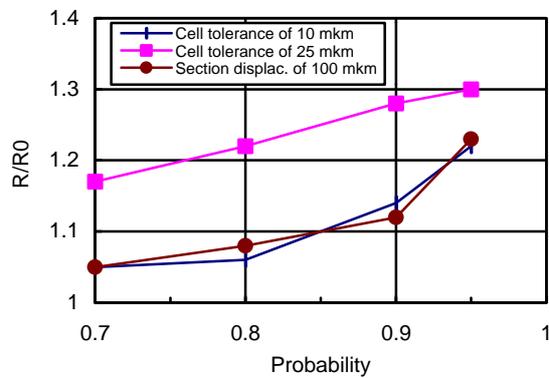

Fig.3. Beam transverse sizes $R/R_0$ (on per-unit basis). Ideal Value equals 1.00

In general it can be stated that $\delta r = 10$ μm are acceptable whereas $\delta r = 25$ μm are dangerous.

In order to suppress parasitic oscillations eight meters long IPHI RFQ cavity must divide on eight one meter long sections with gaps between them. Each even gap is small and its length equals 0.1 mm. Each odd gap is major and its length equals 2.2 mm. The ends of gaps have a curvature of elliptic form.

The effects caused by RF field distortions in intersection gaps have revealed by consideration different RFQ versions that differ by gap number and gap center position. For comparison it were considered versions with gap center placed at cell end (E-versions) and with gap center placed in the point where all field components acting on equilibrium particle are equal to zero (F-versions). It is expected that in the last case beam output parameters will be nearer to nominal ones (without gaps) then in the first case.

The main parameters of output beam are shown in Table 1. Number of gaps and their position (E or F) are used for model version notation.

RF field maximal surface intensity inside gap, $E_g$, RF field maximal surface intensity in the same cell without gap, $E_n$, ratio of the above values and RF field maximal surface intensity in Kilpatrick units ($E_k = 18.43$ ÌV/m) are listed in Table 2.

*Table 2*

| Major Gap | $E_g$ | $E_n$ | $E_g/E_n$ | $E_g/E_k$ |
|---|---|---|---|---|
| 1 (2m) | 33.9 | 30.95 | 1.1 | 1.84 |
| 2 (4m) | 35.0 | 31.9 | 1.1 | 1.9 |
| 3 (6m) | 36.8 | 32.6 | 1.13 | 2.0 |

The main conclusion that can be made on the base of obtained results is the following: presence of gaps with length no more then 2.2 mm and curvature by ellipsoid with semi-axis $a_z = 2$ mm and $a_r = 0.75$ mm does not lead to degradation of output beam parameters if gap center is placed in the point where all field components acting on equilibrium particle are equal to zero (F-versions). In this case RMS emittance growth equals 4 %. In the case when field components acting on equilibrium particle are relatively high (gap center is placed at cell end, F-versions) RMS-emittance growth equals 12 %. In all cases the first gap influence is mainly responsible.

*Table 1*

| Channel Model | Total transm., % | Acc. particle transm, % | Lost particle power, kW | Non accel. particle power, kW | rms-emittance $E_{rx}$, mm·mrad | rms emittance $E_{ry}$, mm·mrad | Total emittance $E_{tx}$, mm·mrad | Total emittance $E_{ty}$, mm·mrad |
|---|---|---|---|---|---|---|---|---|
| No gaps | 98.2 | 95.3 | 0.27 | 0.83 | 0.25 | 0.25 | 2.51 | 2.53 |
| 7E | 96.5 | 92.7 | 2.82 | 1.04 | 0.31 | 0.31 | 4.07 | 3.99 |
| 7F | 97.7 | 94.1 | 0.32 | 1.08 | 0.27 | 0.27 | 3.08 | 3.04 |